\documentclass[12pt]{article}
\usepackage[mathscr]{eucal}
\usepackage[utf8]{inputenc}
\usepackage{feynmp}
\usepackage{tikz-feynman}
\tikzfeynmanset{compat=1.1.0}
\usepackage{slashed}
\usepackage{subcaption}
\usepackage{amssymb,amsmath}

\begin{document}

\begin{titlepage}

\begin{flushright}
arXiv:2210.06668
\end{flushright}
\vskip 2.5cm

\begin{center}
{\Large \bf Aspects of the Equivalence Between the $f^{\mu}$ and $c^{\nu\mu}$\\
Terms in Lorentz-Violating Quantum Field Theory}
\end{center}

\vspace{1ex}

\begin{center}
{\large Sapan Karki and Brett Altschul\footnote{{\tt altschul@mailbox.sc.edu}}}

\vspace{5mm}
{\sl Department of Physics and Astronomy} \\
{\sl University of South Carolina} \\
{\sl Columbia, SC 29208} \\
\end{center}

\vspace{2.5ex}

\medskip

\centerline {\bf Abstract}

\bigskip

It is known that in Lorentz-violating effective field theory, there is a
classical equivalence between certain
coefficients ($c$ and $f$), in spite of the fact that the operators the
two types of coefficients describe appear to have
opposite behaviors under \textbf{CPT}. This paper
is a continuation of previous work extending this equivalence to
the quantum level: generalizing the explicit spinorial point 
transformations that interconvert the $c$ and $f$ terms; demonstrating that the transformations do not
give rise to any additional anomaly terms as the quantum level; and giving explicit prescriptions for
modifying the \textbf{C}, \textbf{P}, and \textbf{T} operators in the $f$ theory, so that they correspond to
the correct interchanges of physical particle states.

\bigskip

\end{titlepage}

\newpage

\section{Introduction}

The standard model is one of the basic pillars of modern physics. It is one of the most thoroughly tested
theories ever devised and has been excellent in explaining the small-scale structure of the universe.
Nevertheless, extending the standard model has been an interesting avenue of research for a long time.
Searches for evidence of whatever exists beyond the standard model could eventually
lead to the discovery of new specific phenomena,
new elementary particles, and perhaps wholly new kinds physics.

One intriguing possibility for new physics beyond the standard model is that the basic symmetries of the theory
might change. In particular, the standard model's (and general relativity's) Lorentz and
\textbf{CPT} invariances
might be merely low-energy approximations, not true in a more basic theory.
The low-energy
effective field theory containing all the new operators which incorporate local, Lorentz-violating
modifications to the physics of known fields is called the standard model extension
(SME)~\cite{ref-kost1,ref-kost2}.
The renormalizable subsector obtained from the SME by removing all the parameters with mass dimension
greater than four has been termed the minimal SME. It has been pointed out by Greenberg~\cite{ref-greenberg}
that \textbf{CPT} violation implies Lorentz violation in a local, stable theory---although Lorentz
violation does not always imply \textbf{CPT} violation. Hence
the action for the minimal SME contains a large number of Lorentz-violating terms,
each of which may or may not be \textbf{CPT}
violating. A rule of thumb for these new structures is that an operator with an even
number of Lorentz indices is \textbf{CPT} even, and one with an odd number of indices is \textbf{CPT} odd.
However, a curious case arises~\cite{ref-altschul8} that the apparently \textbf{CPT}-violating
theory containing a one-index
$f^{\mu}$ term is equivalent to a \textbf{CPT}-preserving theory containing
only a (two-index) $c^{\nu\mu}$ term, by means of a point
transformation of spinor fields. The relation established between the $c$ and $f$ terms
describes an effective $c$ that is of even order in $f$,
which implies that physical effects at first-order in $f$ term are not observable.
Moreover, in interacting theories, these kinds of equivalences can be taken even further;
for example, in the quantum
electrodynamics sector of the SME, an electron $c$ term is completely equivalent to a certain kind of
photon $k_{F}$ term.

The renormalization of the minimal SME, particularly to one-loop order, has been worked on
extensively~\cite{ref-kost3,ref-berr,ref-collad-3,ref-collad-2,ref-collad-1,ref-gomes,ref-anber,ref-ferrero3,
ref-brito1}.
However, even this perturbative renormalization still has not been carried through in all the sectors of
the minimal SME. An example would be that the one-loop renormalization of a Lorentz-violating
gauge theory coupled to charged scalar fields remains incomplete, and hence the
renormalization group (RG) scalings of several Lorentz-violating couplings in the $SU(2)_{L}$
weak gauge sector remain undetermined. Another interesting renormalization issue occurs in the minimal SME
with a Chern-Simons-like term. This term not only breaks Lorentz invariance but also gauge invariance of the
Lagrange density (although the action is gauge invariant), and it was quite controversial as to whether there
could ever be a purely radiatively-generated Chern-Simons term, since it was found that
calculations using different regulators
provided different expressions for the radiative
correction. (See, for example, the discussions in Refs.~\cite{ref-victoria2,ref-altschul37}.)
A similarly curious case was pointed out in Ref.~\cite{ref-karki2}, that the RG scalings of $c$ and
$f$ appeared to be dissimilar, even though the theory with $c$ was known to be equivalent to the $f$ theory
via an appropriate transformation. Fixing the renormalization at second order in $f$ revealed
ambiguities in the $\beta$-functions for $c$ and $f$. However, the RG evolution of the physically
observable quantity $c^{\mu\nu}-\frac{1}{2}f^{\mu}f^{\nu}$ was found to be free of such ambiguities,
up to second order in $f$.

This paper is a continuation of our previous work on the relation between
the SME $c$ and $f$ terms. We shall prsent some results that are relevant to the general structure of the
corresponding $c$ and $f$ theories, with particular emphasis on how those results are related to the
renormalization issues that have previously been discussed.
We observed that a Lagrangian with $f$-type Lorentz violation can be converted into a
Lagrangian with a $c$ instead by means of a point transformation of spinors. In section~\ref{sec-relations}
we generalize the transformation discussed in Ref.~\cite{ref-altschul8}, which showed how to convert a pure-$f$
theory into a pure-$c$ theory. The generalized transformation eliminates $f$ (at all orders)
from a theory that already starts with both $f$ and $c$ terms.

Another closely related question that arises is the
validity of various expressions in the context of a continuation of the
one-loop renormalization to higher orders.
Some of the formulas that we utilized in our $\mathcal{O}(f^{2})$
analysis of the renormalization problem may be modified at higher orders in the Lorentz violation.
Since our analysis used both the Gordon decomposition identity and the closure relations for Dirac spinors,
which do not generally hold without modification in the SME, in calculations beyond second order in $f$
we would need to use the generalized forms of these identities. The generalized expressions will
give rise to extra higher-order terms that we would need to be careful about
accounting for. The ways in which the identities
in question are modified is discussed in section~\ref{sec-identities}.

In section~\ref{sec-quantum}, we turn briefly to the question of whether the classical transformation
that exchanges $f$ for $c$ might become anomalous in a quantum-mechanical context.
Because of operator ordering ambiguities, classical canonical transformations may give rise to additional
terms in the effective potential when a theory is rewritten in new coordinates.
The question naturally arises whether the clean classical equivalence between $c$ and $f$
will also be befouled by an extra potential term at $\mathcal{O}(\hbar)$.

The last major issue discussed in this paper, in section~\ref{sec-discrete}, has to do with charge
conjugation, parity, and time reversal (\textbf{C}, \textbf{P}, and \textbf{T}) properties.
We already know that $c$ is separately \textbf{C} and \textbf{PT} invariant, which makes it \textbf{CPT} even.
However, looking at the usual transformation properties for Dirac bilinears, it appears that  
$f$ must be \textbf{PT} even and \textbf{C} odd, which obviously would make it \textbf{CPT} odd.
If both the actions (either with $c$ or with $f$) represent the same quantum theory,
then there must be some modified \textbf{C}, \textbf{P}, and \textbf{T} operators which will make the
$f$ term overall \textbf{CPT} even. We shall devise a method with which we may convert our usual discrete
spacetime operators that work for the $c$ term into new operators that make $f$
indeed \textbf{CPT} even.

\section{Relations Between Equivalent $c^{\mu\nu}$ and $f^{\mu}$ Terms}

\label{sec-relations}

The transformation of a Lagrange density
\begin{equation}
\mathcal{L}_{f}=\bar{\psi}[i(\gamma^{\mu}+if^{\mu}\gamma_{5})\partial_{\mu}-M]\psi
\end{equation}
with solely an $f$ term into one
\begin{equation}
\mathcal{L}_{c}=\bar{\psi}'[i(\gamma^{\mu}+c^{\nu\mu}\gamma_{\nu})\partial_{\mu}-M]\psi'
\end{equation}
with only a $c$ term has previously been laid out~\cite{ref-altschul8}. The necessary transformations are
\begin{equation}
\label{eq-f-trans}
\psi'=e^{\frac{i}{2}f^{\nu}\gamma_{\nu}\gamma_{5}G(-f^2)}\psi,
\end{equation}
[where $G(\xi)=\frac{1}{\sqrt{\xi}}\tan^{-1}\sqrt{\xi}$ is an analytic function of
its argument $\xi=-f^{2}$], and the corresponding $\bar{\psi}'=(\psi')^{\dagger}\gamma_{0}$.
This converts $\mathcal{L}_{f}$ into the $\mathcal{L}_{c}$ form,
with an effective $c$ term which is given by 
\begin{equation}
\label{eq-ceff-old}
c^{\mu\nu}_{{\rm eff}}=\frac{f^{\mu}f^{\nu}}{f^2}\left(\sqrt{1-f^2}-1\right).
\end{equation}

However, this may be further generalized. (Generalizations to higher-dimensional operators, which resemble the
$c$ and $f$ terms but with additional spacetime derivatives, have already been studied~\cite{ref-schreck3}.)
If we have a Lagrangian with both $c$ and $f$ terms, a relevant
question is whether the theory can be converted into one which only has an effective $c$-type term
or into one with only an effective $f$-type term.
The answer is {\em yes} if the initial $c^{\nu\mu}$ is $c^{\nu\mu}=\alpha v^{\nu}v^{\nu}$, provided that
our $f$ is also in the same direction as the vector $v$, which is the case in many theories, including
single-field bumblebee models of spontaneous Lorentz violation~\cite{ref-kost12}.
A bumblebee models features a dynamical four-vector field which obtains a vacuum expectation value
$v$ through
spontaneous symmetry breaking. So while taking there to be only a single Lorentz-violating spacetime
direction is a significant limitation on the space of possible Lorentz-violating theories we shall
consider, it is not an unnatural or unduly severe limitation.

On the other hand, starting with the most general combination of $f$ and
$c$ terms in the Lagrange density,
\begin{equation}
\label{eq-L-fc}
{\cal L}=\bar{\psi}[i(\gamma^{\mu}+c^{\nu\mu}\gamma_{\nu}+if^{\mu}\gamma_{5})
\partial_{\mu}-M]\psi,
\end{equation}
it is clearly not possible to transfer all the Lorentz violation into the axial vector
$f^{\mu}$ parameter. The two-index tensor $c^{\nu\mu}$ contains nine physically
observable parameters. [These are not precisely identical with the traceless, symmetry part
$c^{(\nu\mu)}=c^{\nu\mu}+c^{\mu\nu}$ of the tensor, when considered beyond leading order.
In fact, the nine physical coefficients are the encapsulated by the traceless part of
the manifestly symmetric
tensor $c^{(\nu\mu)}+c^{\alpha\nu}c_{\alpha}\,^{\mu}$.] The four-component $f^{\mu}$ cannot
possibly parameterize the full texture of the nine-dimensional space of $c^{\nu\mu}$ parameters.

Yet if all Lorentz violation in a model is described by a single underlying vectorial
quantity, then the $c$ and $f$ theories are, as noted above, completely equivalent. 
One way to see how this works is via the following iterative approach.
After the transformation of the Dirac spinor $\psi$ into $\psi'$ using relation~(\ref{eq-f-trans})
we get a Lagrange density with two kinds of Lorentz-violating terms:
a term of the form $ic^{\nu\mu}f_{\nu}\gamma_{5}$, which is a $f$-type term and is of order
$\mathcal{O}(v^{3})$;
and an effective $c$-type term  that looks like $c^{\nu\mu}-\frac{1}{2}f^{\nu}f^{\mu}$ to $\mathcal{O}(v^{2})$.
If we repeat the transformation, with the new effective $f$-type term (the $ic^{\nu\mu}f_{\nu}\gamma_{5}$),
this $\mathcal{O}(v^{3})$ term can be absorbed into a $\mathcal{O}(v^{6})$ $c$-type contribution, although we
are again left with a $f$-type term, $i(c^{\nu\mu}-\frac{1}{2}f^{\nu}f^{\mu})c_{\nu\alpha}f^{\alpha}\gamma_{5}$,
up to the order $O(v^{5})$. This process can be continued indefinitely, and we can get rid of the $f$
term totally, working
order by order; the sequence of transformed Lagrangians will obviously converge if $v$ is a sufficiently small
parameter (as we expect a Lorentz-violating vacuum expectation to be).

We can also get the exact transformation in a form similar to~(\ref{eq-f-trans}) directly.
Assuming that there exists such a transformation which can get rid of the $f$-type term in the Lagrangian
(as guaranteed by the preceding iterative argument), it should take the same general form
as~(\ref{eq-f-trans}),
\begin{equation}
\label{eq-theta-trans}
\psi'=A\psi=e^{\frac{i}{2}f^{\nu}\gamma_{\nu}\gamma_{5}\frac{\theta}{\sqrt{f^2}}}\psi,
\end{equation}
albeit with an as yet unknown parameter $\theta$. The crucial matrix $A$ may be expanded directly as
\begin{equation}
\label{eq-A}
A=\cosh\frac{\theta}{2}+i\frac{f^{\nu}}{\sqrt{f^{2}}}\sinh\frac{\theta}{2}\gamma_{\nu}\gamma_{5}.
\end{equation}
Since (\ref{eq-theta-trans}) is equivalent to a change
in the Dirac matrices instead of the fermion field,
the transformation of the matrix $\gamma^{\mu}$ under it is
\begin{equation}
\label{eq-gamma-trans}
\left(\gamma^{\mu}\right)'=e^{\frac{i}{2}f^{\nu}\gamma_{\nu}\gamma_{5}\frac{\theta}{\sqrt{f^2}}}
\gamma^{\mu}e^{-\frac{i}{2}f^{\nu}\gamma_{\nu}\gamma_{5}\frac{\theta}{\sqrt{f^2}}}
=\gamma^{\mu}-\sinh\theta\frac{if^{\mu}}{\sqrt{f^2}}\gamma_{5}
+(\cosh\theta-1)\frac{f^{\mu}f^{\nu}}{f^{2}}\gamma_{\nu}.
\end{equation}
This gives the transformation of the $c^{\nu\mu}\gamma_{\nu}=\alpha f^{\nu}f^{\mu}\gamma_{\nu}$ term
in the Dirac Lagrangian. (We now set $v^{\mu}=f^{\mu}$ for simplicity
of notation, since $v$ is no longer needed as a power counting parameter.) The transformation of the
$if^{\mu}\gamma_{5}$ term in (\ref{eq-L-fc}) in  may be computed similarly.
After the transformation we require that there be no $f^{\mu}$ term remaining in the transformed
$\mathcal{L}'$; using this condition we may compute the angle $\theta$ and find
\begin{equation}
\label{eq-theta}
\tanh\theta=\frac{\sqrt{f^2}}{1+\alpha f^2},
\end{equation}
observing that $\alpha=0$ reproduces the transformation~(\ref{eq-f-trans}).
With this $\theta$, we can then evaluate the final effective $c$-type term,
\begin{equation}
\label{eq-ceff}
c^{\nu\mu}_{{\rm eff}}=-\frac{f^{\nu}f^{\mu}}{f^{2}}\left[1-\sqrt{(1+\alpha f^2)^2-f^2}\right].
\end{equation}
Naturally, this 
correctly reproduces the special case~(\ref{eq-ceff-old}) if $\alpha=0$.
It is also clearly correct in the opposite limit of no initial $f$ term (so $f\rightarrow 0$,
which corresponds to
$\alpha\rightarrow\infty$), yielding the finite $c^{\nu\mu}_{{\rm eff}}=c^{\nu\mu}=\alpha f^{\nu}f^{\mu}$.

\section{Spinor and Propagator Identities}

\label{sec-identities}

In our paper~\cite{ref-karki2},
we converted two $f$ vertices in an explicit momentum-space
propagator to a single effective $c$ vertex in the
propagator; the effective $c^{\mu\nu}$ term was given by 
\begin{equation}
c^{\mu\nu}_{{\rm eff}}=-\frac{1}{2}f^{\mu}f^{\nu}.
\end{equation}
This again reaffirms that the relation (\ref{eq-ceff-old}) is indeed correct, at least at the lowest order.
We used several approximations while deriving this result which did not affect the results up to
$\mathcal{O}(f^{2})$. However, when computing third-order radiative corrections, it may be necessary
to include the neglected terms. Therefore, let us consider more carefully the effects of two $f$ vertex
insertions into the fermion propagator.

The contribution with these insertions is equal to 
\begin{equation}
\label{eq-2f-prop}
S_{ff}=S(p)(-f^{\mu}\gamma_{5}p_{\mu})S(p)(-f^{\nu}\gamma_{5}p_{\nu})S(p),
\end{equation}
where $S(p)=\frac{i}{\slashed{p}-m}$ is the unmodified propagator. Upon straightforward rearrangement
of the Dirac matrices, (\ref{eq-2f-prop}) gives
\begin{eqnarray}
S_{ff}=S(p)\frac{1}{(p^2-m^2)}f^{\mu}f^{\nu}p_{\mu}p_{\nu}.
\end{eqnarray}
The trick to simplifying this further runs as follows. Use the relation
$\slashed{p}+m=\sum u(p)\bar{u}(p)$ in the
numerator of the remaining $S(p)$. Then use the
Gordon decomposition to change one $p_{\alpha}$ to $m\gamma_{\alpha}$ by sandwiching it between the
$u(p)$ and $\bar{u}(p)$. Then finally the mass term can be changed to one-half of $\slashed{p}$.
After these steps, the remaining expression looks like a propagator with a $c$-type vertex insertion.
However, upon closer inspection, it turns out that each of these step
may potentially have additional corrections at higher
orders in $f$; hence a more careful analysis is required if we are to make any statements about higher
order corrections. We will therefore look at the generalizations of two key identities that are
frequently used in Dirac algebra.

The modified relation (the momentum-space Dirac equation) satisfied by the spinors with the SME $c$ and
$f$ terms is 
\begin{equation}
\label{eq-mod-Dirac}
\left(\slashed{{p}}+m+c^{\nu\mu}\gamma_{\nu}p_{\mu}+if^{\mu}\gamma_{5}p_{\mu}\right)u(p)=0.
\end{equation}
The modified Gordon identity for these spinors is consequently
\begin{equation}
\bar{u}(p')\gamma^{\mu}u(p)=\bar{u}(p')\left[\frac{(p+p')^{\mu}}{2m}+i(\sigma^{\mu\nu}
-ic^{\rho\nu}\gamma_{\rho}\gamma^{\mu}+f^{\nu}\gamma_{5}\gamma^{\mu})\frac{(p'-p)_{\nu}}{2m}
+\frac{c^{\mu\nu}p_{\nu}}{m}\right]u(p).
\end{equation}
This means that when $p'=p$ and if we consider a theory with only the $f$ term, we can still exchange
$\gamma^{\mu}$ for $mp^{\mu}$ as we did our previous calculation of $S_{ff}$; that trick
still works and is fine.

The second very useful quantity is the spin-summed outer product
$\sum u(p)\bar{u}(p)$ in the theory with just a $f$ term.
To calculate this, we can evaluate the expression in
the fermion rest frame and then boost the expression (taking care to boost the Lorentz-violating backgrounds
as well). The result is 
\begin{equation}
\sum u(p)\bar{u}(p)=\slashed{{p}}+m+if^{\mu}\gamma_{5}p_{\mu}.
\end{equation}
This relation can be used, along with the modified Gordon decomposition, to simplify calculations at
higher orders in $f$.

However, we need to be careful when we include a $c$ term in the starting Lagrangian along with an
equivalent $f$,
as we can see that we get an additional effective $c$ term which is second order in $f$, and
according to (\ref{eq-ceff}), the effective $c$-type terms are not simply additive.
Calculations may be carried through order by order in $v$, keeping $c$ and $f$ contributions of equivalent
orders. Working in this fashion, all contributions at odd orders in $v$ should cancel in physical
observables. Alternatively,
this problem can be circumvented if we convert the
starting theory into an equivalent theory which only contains
the $c$ or $f$ term and work entirely in that theory, transforming back to the original only that the very end
of calculation.

%

\section{Quantum Considerations}

\label{sec-quantum}

Another issue is whether the relation (\ref{eq-ceff}) is  exact in quantum theory.
Relations that hold at the classical
level are sometimes no longer valid after the quantization of a theory. The problem typically comes form the
fact that several quantum theories may have the same classical limit, due to operator ordering ambiguities.
For example, it has been pointed out~\cite{ref-gulyaev,ref-christ1} 
that in the path integral formalism, when we perform
a change of coordinates from Cartesian to polar, we get additional (effective) potential energy terms,
which compensate for the fact that the new canonical coordinates are not linear in the previous
coordinates. The rotation symmetry (corresponding to the conserved angular momentum) gives rise
to a zero-energy mode, and although we can get rid of this zero mode, we obtain an additional contribution of
$\frac{L^{2}}{2mr^2}$ to the effective potential---the classical centrifugal term. Another important example
would be the case of kink solitons in 1+1 dimensions~\cite{ref-christ2}.
The kinks, which are topological solutions
of the classical equations of motion for a scalar field, have a translation
symmetry. A consequence of this zero-mode symmetry is that perturbative computations
diverge while computing the higher order corrections, as the corrections have $\omega_{0}$ (the zero-mode
frequency) in the denominator~\cite{ref-jevicki}. We can get rid of this problem by performing a point
transformation and computing the path integral in suitable coordinates, so that we get
rid of the troublesome zero-mode coordinate; but again the transformation gives rise to additional
new terms in the effective potential.

If the transformation that takes a $f$ theory to a $c$ theory is associated with an additional
effective potential term
at $\mathcal{O}(\hbar)$, then it could  give further corrections to the theory's $c_{{\rm eff}}$.
This would mean that the renormalization process
at higher orders might behave differently from what we have
already computed at the lowest nontrivial orders.

The exact form of the additional potential accompanying a point transformation is (including the
explicit loop-counting parameter $\hbar$)~\cite{ref-gervais},
\begin{equation}
\label{eq-Vprime}
\Delta V'(\hat{Q})=\frac{1}{4}\hbar\left(\frac{1}{2}g^{ij}(\hat{Q}),_{ij}-2g(\hat{Q})^{-\frac{1}{4}}
\left\{g^{ij}(\hat{Q})\left[g^{\frac{1}{2}}g(\hat{Q})^{-\frac{1}{4}}\right],_{i}\right\}_{j}\right),
\end{equation}
where $Q$ is the new coordinate variable; its canonically conjugate momentum would be $P$.
The new metric $g$ for the revised coordinate space is given by
\begin{equation}
\label{eq-g-metric}
g_{ij}=\Sigma_{a} F^{a}(Q),_{i}F^{a}(Q),_{j};
\end{equation}
with the commas in (\ref{eq-Vprime}) and (\ref{eq-g-metric}) indicating differentiation, and
$F$ giving the relationship between the old canonically conjugate variables $q$ are $p$ and the new $Q$
and $P$, according to
\begin{equation}
q^{a}= F^{a}(Q) \quad p^{a}= F^{a},_{i}(Q)g^{ij}P_{j}.
\end{equation}

The Dirac Lagrangian can be cast in the Hamiltonian form with momenta and coordinates
$\psi^{\dagger}$ and $\psi$ corresponding to $p$ and $q$, respectively (at least in flat
Minkowski spacetime).
The relation (\ref{eq-Vprime}) is formulated for a bosonic system. For the fermionic fields, results are
analogous~\cite{ref-oconnor}, with the quantum correction to a canonical point transformation originating
in two ways: either due to the nonlinearity of the transformation, or due to a anomalous Jacobian
for the transformation. We shall discuss each of these in turn.

If we perform the change of canonical coordinates as suggested by the relation (\ref{eq-f-trans})
then the metric $g$ appearing in
(\ref{eq-g-metric}) will be a constant matrix, dependent only on the $\gamma$-matrices and $f$---not
on position or field strength $\psi$. It is then
immediately apparent that the additional potential term given
analogously to (\ref{eq-Vprime}) is identically zero.
Another, perhaps easier, way to see this is by observing the fact that the additional potential terms arise
only when there is a nonlinear change in the dynamical variables, which does not happen with
(\ref{eq-f-trans}) or (\ref{eq-theta-trans}). Hence it turns out that the formula (\ref{eq-ceff}) must be
stable even with the inclusion of quantum corrections.

The equivalence between $f$ and $c$ is obviously most interesting if
the equivalence persists even in the presence of interactions, and this has guided our
choice of methods for demonstrating the equivalence. In this connection, we should note that
there is no chiral anomaly associated with the transformation $A$, even if a gauge interaction
is added to the action. The gauge invariant regulation of the fermion measure leads to a nontrivial
Jacobian $\mathcal{J}$ exactly if
\begin{equation}
-i\log\mathcal{J}=\lim_{\Lambda\rightarrow\infty} \int d^{4}x\,\langle x|\operatorname{tr}\left[(\log A)
e^{(i\slashed{D})^{2}/\Lambda^{2}}\right]|x\rangle
\end{equation}
is nonzero~\cite{ref-fujikawa}. However, since $\log A$ is proportional to $\gamma_{\nu}\gamma_{5}$,
the expansion of the exponential regulator always produces traces of odd numbers of basic $\gamma$-matrices,
which uniformly vanish.

Moreover,
the fact that the effects of (\ref{eq-theta-trans}) do not produce any additional potential terms can also
be seen in yet another way---via an analogy to a slightly simpler transformation that is known not
to be associated with complications of this sort.
In the path integral formalism,  $\Delta V'$ arises out of the
nontrivial transformation of the path integral measure.
When moving from integrals over Grassmann variables to path integrals over the Grassmann-valued fields,
it is standard to change the measure from $\mathcal{D}\psi^{*}\,\mathcal{D}\psi$ to
$\mathcal{D}\bar{\psi}\,\mathcal{D}\psi$, often without even mentioning the change, even in
pedagogical treatments. The measures are, of course, completely equivalent, but there is a unitary
transformation that effects the change, $\bar{\psi}_{a}=\left(\gamma_{0}\right)_{ab}\psi^{*}_{b}$.
For the reasons discussed in the previous two paragraphs, this transformation to new dynamical variables
does not introduce any additional potentials; the transformation simply reshuffles the components of one
of the Dirac fields, in the same fashion at all spacetime locations.  This is exactly what
(\ref{eq-theta-trans}) represents as well---a position- and momentum-independent similarity transformation
of the components of the Dirac spinor fields. Just as $\mathcal{D}\psi^{*}\,\mathcal{D}\psi\rightarrow
\mathcal{D}\bar{\psi}\,\mathcal{D}\psi$ is an uncomplicated transformation of the measure, so is
$\mathcal{D}\bar{\psi}\,\mathcal{D}\psi\rightarrow\mathcal{D}\bar{\psi}'\,\mathcal{D}\psi'$.

\section{Discrete Symmetries}

\label{sec-discrete}

The last major topic we shall address is the discrete symmetries of the $f$ theory. This is an issue that
has the potential to
cause a fair amount of confusion, since---to naive appearances---a theory with only a $c$-type term must be
even under \textbf{C} and \textbf{PT}, while a theory with just a $f$ has different behavior.
That might make it difficult to see how theories with $c$ and $f$ could be equivalent.
Indeed, a Lagrangian with just $c^{00}$ or $c^{jk}$ terms manifestly corresponds to a
\textbf{C}-even, \textbf{P}-even, \textbf{T}-even theory. However, it appears that $f^{0}$
is actually separately odd under \textbf{C}, \textbf{P}, and \textbf{T}, which presents a puzzle.

The solution to the puzzle  is that the \textbf{C}, \textbf{P}, and \textbf{T} operators have to
be modified if the theory contains an $f$ term. The usual forms of these operators are derived from the
fact that they leave the Lagrangians for certain theories (such as the free Dirac theory, or its
extension to quantum electrodynamics) invariant. However, while the way the symmetries act on
coordinates is fixed by the fact that certain coordinates must be inverted or preserved,
the way the operators act in the four-dimensional Dirac spinor space may be adjusted and is not so
strictly fixed.
It is familiar, for example, that the phases associated with 
the discrete operators cannot be uniquely determined, or that each scalar field in a theory may
be assigned an even or odd intrinsic parity. In the $f$ theory, the discrete operators are modified
in a much more profound way, but they are still valid representations of the \textbf{CPT} algebra; and
as long as a representation exists that leaves the action of the
theory invariant, the physical observables of the
theory will be likewise invariant.

Once this is recognized, it is not too challenging to determine the forms of the modified operators,
since it is already known how to transform the $f$ theory into a theory with a $c_{{\rm eff}}$.
For simplicity, we shall consider a theory with only a $f$, but the generalization to a theory with
both types of coefficients is straightforward---simply using the transformation (\ref{eq-theta-trans})
instead of (\ref{eq-f-trans}) in what follows.

\subsection{Parity and Reflections}

There are other minimal SME terms besides $f^{0}$
that share the property of being odd under the standard \textbf{C}, \textbf{P}, and \textbf{T}
operations---spatial $a^{j}$ and $e^{j}$ terms, for instance.
However, there are additional, less-discussed discrete symmetries that distinguish these operators.
The parity operator is defined as inverting all three spatial coordinates, taking
$\vec{x}\rightarrow-\vec{x}$. However, this \textbf{P} may be broken down into the product of
three separate
spatial reflections, $\textbf{P}=\textbf{R}_{1}\textbf{R}_{2}\textbf{R}_{3}$, where $\textbf{R}_{j}$
inverts just one of the coordinates, taking $x_{j}
\rightarrow-x_{j}$ and leaving the two orthogonal coordinates unaffected. In Lorentz-invariant
theories, little distinction is made between the behavior of a quantities under \textbf{P} and
$\textbf{R}_{j}$, because \textbf{P} can also be written as $\textbf{R}_{j}$ followed by a $\pi$ rotation
around the $x_{j}$-axis. However, in the absence of rotation symmetry, the behavior of
observables under \textbf{P} and $\textbf{R}_{j}$ need not be the same.

While particular spacelike components
$a^{j}$ and $e^{j}$ are odd under $\textbf{R}_{j}$, which inverts the axis along which those coefficients point,
they are even under reflections along the two perpendicular axes.
However, the behavior of $f^{0}$ different; using the standard reflection operator
$\textbf{R}_{j}$ (which acts on the Dirac spinor $\psi$ by the matrix
$S_{R_{j}}=i\gamma^{j}\gamma_{5}$), the $f^{0}$ term appears to be odd under each of the 
$\textbf{R}_{j}$ separately (or, indeed
under any reflection whatsoever, including ones along oblique directions).
This reflection behavior is unique in the minimal SME; it is shared by no other term.

Considering the theory with a $f$ term,
the Lagrange density contains the Dirac matric operator $\gamma^{\mu}+if^{\mu}\gamma_{5}$ sandwiched between
the fermion fields. Converting this theory into the theory with the $c_{{\rm eff}}$ term
is equivalent to using the conjugation operation
\begin{equation}
\label{eq-A-def}
\gamma^{\mu}+if^{\mu}\gamma_{5}\rightarrow
A(\gamma^{\mu}+if^{\mu}\gamma_{5})A^{-1}\equiv
e^{\frac{i}{2}f^{\nu}\gamma_{\nu}\gamma_{5}\frac{\theta}{\sqrt{f^{2}}}}
(\gamma^{\mu}+if^{\mu}\gamma_{5}) e^{-\frac{i}{2}f^{\nu}\gamma_{\nu}\gamma_{5}\frac{\theta}{\sqrt{f^{2}}}}
=\gamma^{\mu}+c_{{\rm eff}}^{\nu\mu}\gamma_{\nu}
\end{equation}
(with $\theta=\tanh^{-1}\sqrt{f^{2}})$ on the kinetic operator in spinor space. [Note that the $A$ that
appears on left in $A(\gamma^{\mu}+if^{\mu}\gamma_{5})A^{-1}$ actually arises---due to the presence
of $\bar{\psi}=\psi^{\dagger}\gamma^{0}$ on the left-most end of the operator product in the action---as the
inverse of $\overline{A^{-1}}=\gamma^{0}(A^{-1})^{\dagger}\gamma^{0}=A^{-1}$.]
Under this conjugation, the $if^{\mu}\gamma_{5}$ mixes with the $\gamma^{\mu}$ to produce the
linear combination $\gamma^{\mu}+c_{{\rm eff}}^{\nu\mu}\gamma_{\nu}$ of $\gamma$-matrices.
To understand the physical behavior of the $f$ theory under a spacetime transformation, we should look at
how the full kinetic sector of the theory behaves---considering the \textbf{C}, \textbf{P}, and \textbf{T}
properties of $\gamma^{\mu}+if^{\mu}\gamma_{5}$ as a single block, rather than
looking at the properties of $f$ or $c$ terms on their own.
With this in mind, we consider a discrete symmetry
(involution) operator $\textbf{X}'$ that acts on $\psi'$, where
$\psi'$ is the fermion field in a theory transformed to have
solely a $c$-type term in Lagrangian. The action of conjugation by $\textbf{X}'$ on the field
operator is represented
by a linear transformation with matrix $S_{X}'$, according to $\textbf{X}'\psi'\textbf{X}'=S_{X}'\psi'$.

Acting with $S_{X}'$ on $\psi'$ is equivalent to conjugating $\gamma^{\mu}+c^{\mu\nu}\gamma_{\nu}$ by
$S_{X}'$, and if
$\textbf{X}$ is a symmetry of the action, then the conjugation operation must have an eigenvalue $(-1)^{X}$,
\begin{equation}
\label{eq-SXprime}
(-1)^{X}(\gamma^{\mu}+c^{\nu\mu}\gamma_{\nu})=S_{X}'(\gamma^{\mu}+c^{\nu\mu}\gamma_{\nu})S_{X}'.
\end{equation}
To find $S_{X}$, the matrix representation of how \textbf{X} acts in the untransformed theory with $f$,
we may simply insert factors of $A$ and $A^{-1}$ into (\ref{eq-SXprime}) and use (\ref{eq-A-def})
on each side,
\begin{eqnarray}
(-1)^{X}A^{-1}(\gamma^{\mu}+c^{\mu\nu}\gamma_{\nu})A & = & A^{-1}S_{X}'
(\gamma^{\mu}+c^{\mu\nu}\gamma_{\nu})S_{X}'A \\
(-1)^{X}(\gamma^{\mu}+if^{\mu}\gamma_{5}) & = & A^{-1}S_{X}'A(\gamma^{\mu}+if^{\mu}\gamma_{5})A^{-1}S_{X}'A,
\end{eqnarray}
then read off the expression $S_{X}=A^{-1}S_{X}'A$.

In the $\psi'$ theory, $\gamma^{0}$ acts as the parity transformation matrix $S_{P}'$. It is straightforward
to find the equivalent operation of \textbf{P} in the $\psi$ theory that contains $f$ but no $c$,
\begin{eqnarray}
\textbf{P}\psi\textbf{P}=S_{P}\psi(t,-\vec{x})
& = & e^{-\frac{i}{2}f^{\nu}\gamma_{\nu}\gamma_{5}\frac{\theta}{\sqrt{f^{2}}}}
\gamma^{0}e^{\frac{i}{2}f^{\nu}\gamma_{\nu}\gamma_{5}\frac{\theta}{\sqrt{f^{2}}}}\psi(t,-\vec{x}) \\
\label{eq-SP}
& = & \left[\gamma^{0}+i\frac{f^{0}}{\sqrt{1-f^{2}}}\gamma_{5}
+\frac{\left(1-\sqrt{1-f^{2}}\right)f^{0}f^{\nu}}{f^2\sqrt{1-f^{2}}}\gamma_{\nu}
\right]\psi(t,-\vec{x}),\quad
\end{eqnarray}
much like in (\ref{eq-gamma-trans}). In spite of the presence of a $f^{2}$ factor in the denominator
of the last bracketed term of $S_{P}$ in (\ref{eq-SP}),
the whole expression is regular as $f^{2}\rightarrow 0$, since
the numerator contains $1-\sqrt{1-f^{2}}$, which is approximately $\frac{1}{2}f^{2}$ in that limit.

Note that $S_{P}$ in (\ref{eq-SP}) is not Lorentz invariant, but instead it has, just like $S_{P}'=\gamma^{0}$, 
exactly one free time index in each term. Moreover, if $f^{0}=0$, then the $S_{P}$ matrix is
not modified, which is correct, since theories in which either $f$ is purely spacelike or $c$ has only
space-space (that is, $c^{jk}$) components, there is no negative sign under the action of $S_{P}'$
on $\psi$. For a less trivial example, let us consider the next-simplest case, in which
$f^{\mu}=(f^{0},0,0,0)$ is purely timelike.
Then the new parity matrix is
\begin{equation}
S_{P}=\frac{1}{\sqrt{1-(f^{0})^2}}\gamma^{0}+i\frac{f^{0}}{\sqrt{1-(f^{0})^2}}\gamma_{5}.
\end{equation}
This is directly proportional to the matrix that multiplies the time derivative $\partial_{0}$
in the Dirac Lagrange density
$\mathcal{L}_{f}$, and it also anticommutes with all the spatial $\gamma^{j}$ matrices. Thus
one can easily see that the $\gamma^{\mu}\partial_{\mu}+if^{0}\gamma_{5}\partial_{0}$ term in the Lagrange
density is even under the action of
the conjugation by this modified parity matrix. This makes sense because in this theory $c_{{\rm eff}}$
only has a $c^{00}$ component, and the term $c^{00}\gamma_{0}\partial_{0}$ is also even under parity.

The same method may be applied to find the correct reflection operators in the $f$ theory. For
example, for inversion of the $x_{1}$-direction, the new $x_{1}$-axis reflection operator
$S_{R_{1}}$ is, following the above line of argument,
\begin{eqnarray}
S_{R_{1}} & = & A^{-1}S_{R_{1}}' A = A^{-1}i\gamma^{1}\gamma_{5}A \\
& = & \frac{i}{\sqrt{1-f^{2}}}\gamma^{1}\gamma_{5}+\frac{f^1}{\sqrt{1-f^2}}I+
\frac{f^{\nu}}{\sqrt{1-f^{2}}}\gamma^{1}\gamma_{\nu}
+\frac{i\left(1-\sqrt{1-f^{2}}\right)f^{1}f^{\nu}}{f^{2}\sqrt{1-f^{2}}}\gamma_{\nu}\gamma_{5}\quad
\end{eqnarray}
(where $I$ is the identity matrix in $4\times 4$ spinor space).
The results for other spatial reflections are obviously analogous.

An interesting corollary of the nonlinearity of the relationship between $f$ and $c_{{\rm eff}}$ is that
the individual components of $f$ do not possess well-defined parities. The parity transformation just
of the Dirac matrices in the $c$-type fermion kinetic term is
\begin{equation}
\textbf{P}'(\gamma^{\mu}+c^{\mu\nu}\gamma_{\nu})\textbf{P}'=
(-1)^{\mu}\gamma^{\mu}+(-1)^{\nu}c^{\nu\mu}\gamma_{\nu},
\end{equation}
where $(-1)^{\mu}$ is $-1$ if $\mu$ is any spacelike index $j$. Including the derivative term
\begin{equation}
\textbf{P}'(\gamma^{\mu}+c^{\mu\nu}\gamma_{\nu})\textbf{P}'\textbf{P}'\partial_{\mu}\textbf{P}'=
\gamma^{\mu}+(-1)^{\nu}(-1)^{\mu}c^{\nu\mu}\gamma_{\nu}.
\end{equation}
Different components of $c$ evidently have different behavior under $\textbf{P}'$; in particular, while
$c^{00}$ and $c^{jk}$ terms are even, the mixed time-space term $c^{0j}$ is $\textbf{P}'$ odd. (The individual
$c^{jk}$ terms can similarly be further classified by their behavior under individual reflections.)
This means that
a $f$ theory with either a purely timelike or purely spacelike $f$ is even under \textbf{P}. However, if
there are both a nonzero $f^{0}$ and nonzero $f^{j}$, then the theory includes physical parity-violating
effects. So the transformation of a term in the $f$ theory into an equivalent $c$-type term
does not give nice \textbf{P} (or, in a similar fashion, \textbf{T}) eigenvalues in general.
However, since every single $c$ term in $\mathcal{L}_{c}$ is symmetric under $\textbf{P}'\textbf{T}'$ and
$\textbf{C}'$ separately,
we expect that every $f$ term should likewise be even under the
modified action of \textbf{P}\textbf{T}, \textbf{C}, and
thus \textbf{C}\textbf{P}\textbf{T} operators.

\subsection{Charge Conjugation}

The purely spatial transformations \textbf{P} and $\textbf{R}_{j}$ are the simplest of the the discrete
symmetries to work with. Like purely spatial rotations, they are represented by unitary matrices that
act on the spinor structure of $\psi$. The remaining operations, \textbf{C} and \textbf{T}, are potentially
trickier, since there are complex conjugations involved. For example, because \textbf{C} exchanges particle
and antiparticle states, $\textbf{C}\psi\textbf{C}=S_{C}\psi^{*}$. (Note that another convention for
the definition of $S_{C}$ also exists, differing from ours by a factor of $\gamma^{0}$.)

The requirement on the $S_{C}'$ matrix (which implements charge conjugation in the standard Dirac theory,
or the theory with only $c$-type Lorentz violation)
is that it should satisfy $S_{C}'\gamma^{\mu}=-(\gamma^{\mu})^{*}S_{C}'$. This ensures that $\textbf{C}'$
interchanges fermion and antifermion creation (and, separately, annihilation) operators. The analogous
condition in the $f$-type theory is that
\begin{equation}
\label{eq-SC-with-f}
S_{C}(\gamma^{\mu}+if^{\mu}\gamma_{5})=-(\gamma^{\mu}+if^{\mu}\gamma_{5})^{*}S_{C}.
\end{equation}
In both the Dirac and Weyl representations of the Dirac algebra (which differ only by the interchange of
$\gamma^{0}$ and $-\gamma_{5}$), $\gamma^{0}$, $\gamma^{1}$, and $\gamma^{3}$ are real, so $S_{C}'$
is proportional to the imaginary $\gamma^{2}$: $S_{C}'=-i\gamma^{2}$ with a standard choice of phase.
[The alternate convention for $S_{C}$ mentioned above is
based on the fact that $\textbf{C}'\psi'\textbf{C}'=-i(\bar{\psi}'\gamma^{0}\gamma^{2})^{T}$ and
$\textbf{C}'\bar{\psi}'\textbf{C}'=(-i\gamma^{0}\gamma^{2}\psi')^{T}$ in the Dirac and Weyl representations.]
To find a matrix $S_{C}$ that satisfies (\ref{eq-SC-with-f}), we can actually just transform $S_{C}'$
in much the same way as $S_{P}'$---just taking account of the extra complex conjugation in 
$\textbf{C}\psi\textbf{C}=S_{C}\psi^{*}$. Doing this, we have
\begin{equation}
\textbf{C}\psi\textbf{C}=S_{C}\psi^{*}=A^{T}S_{C}'A^{*}\psi^{*}.
\end{equation}
[Note that while $A$ can expressed in the representation-independent form (\ref{eq-A}), the matrices
$A^{*}$ and $A^{T}$ cannot, because of the way they depend on which of the representation
matrices are imaginary.]
As an example, we consider again the theory in which $f$ is purely timelike, so that (\ref{eq-A}) gives
\begin{equation}
A=\cosh\frac{\theta}{2}+i\frac{f^{0}}{|f^{0}|}\sinh\frac{\theta}{2}\gamma^{0}\gamma_{5}.
\end{equation}
Applying this $A$, along with (\ref{eq-theta}) with $\alpha=0$, the expression for the charge conjugation
matrix becomes
\begin{equation}
S_{C}=-\frac{i}{\sqrt{1-(f^{0})^{2}}}\gamma^{2}+\frac{f^{0}}{\sqrt{1-(f^{0})^{2}}}
\gamma^{0}\gamma^{2}\gamma_{5},
\end{equation}
which can be checked to satisfy (\ref{eq-SC-with-f}).

In the less-used Majorana representation, all four Dirac $\gamma^{\mu}$ matrices are purely imaginary,
obeying $\gamma_{\mu}^{T}=-(-1)^{\mu}\gamma_{\mu}$ and $\gamma_{5}^{T}=-\gamma_{5}$.
The usual charge conjugation matrix in this representation is $S_{C}'=-iI$. In fact, this is
not actually that puzzling, since if all four $\gamma^{\mu}$ are imaginary, then
$S_{C}'\gamma^{\mu}=-(\gamma^{\mu})^{*}S_{C}'$ just requires that $S_{C}'$ commute with them all; and by
Schur's Lemma, only a multiple of the identity has this property. The simplicity of $S_{C}'$ makes
the evaluation of $S_{C}$ also quite simple (even for arbitrary $f$),
\begin{equation}
S_{C}=A^{T}(-i)A^{*}=-i(A^{*})^{2}=-\frac{i}{\sqrt{1-f^{2}}}I
-\frac{f^{\nu}}{\sqrt{1-f^{2}}}\gamma_{\nu}\gamma_{5}.
\end{equation}

\subsection{Time Reversal}

Finally, let $\textbf{T}$ be the time reversal operator and $S_{T}$ the corresponding matrix.
Attention has to be paid to the fact that time reversal is anti-unitary.
The additional anti-unitarity means that when the matrix is transformed to a different
basis, the $A^{-1}$ on the left (as in $S_{X}=A^{-1}S_{X}'A$) must be complex conjugated.
The time reversal matrix for the Lagrangian with $c$ is $S_{T}'=\gamma^{1}\gamma^{3}$ in the Dirac and
Weyl representations.
A useful identity in these representation is $(A^{-1})^{*}=-\gamma^{2}A^{-1}\gamma^{2}$, so
it turns out that the matrix by which \textbf{T} operates in the $f$ theory is
\begin{eqnarray}
\textbf{T}\psi\textbf{T}&=& (A^{-1})^{*}\gamma^{1}\gamma^{3}A \psi(-t,\vec{x}) \\
& = & \left[-\frac{i}{\sqrt{1-f^{2}}}\gamma^{1}\gamma^{3}+
\frac{f^{\nu}}{\sqrt{1-f^{2}}}\gamma^{1}\gamma^{3}\gamma_{\nu}\gamma_{5}
+\frac{f^{0}}{\sqrt{1-f^2}}\gamma^{2}\right. \nonumber\\
& & \left.+i\frac{\left(1-\sqrt{1-f^{2}}\right)f^{0}f^{\nu}}{f^{2}\sqrt{1-f^{2}}}
\gamma^{2}\gamma_{\nu}\gamma_{5}\right]\psi(-t,\vec{x}) .
\end{eqnarray}

An alternative guiding principle for
verifying the action of \textbf{T}
is that the combination \textbf{TP} of operators
must have the same eigenvalue for its conjugation action on $\gamma^{\mu}+if^{\mu}\gamma_{5}$ as
$\textbf{T}'\textbf{P}'$ has on $\gamma^{\mu}+c^{\mu\nu}\gamma_{\nu}$.
The combined action of both operators is
\begin{eqnarray}
\textbf{TP}\psi\textbf{PT} & = & (A^{-1})^{*}\gamma^{0}\gamma^{1}\gamma^{3}A\psi(-t,-\vec{x}) \\
& = & \left(\frac{i}{\sqrt{1-f^{2}}}\gamma^{2}\gamma_{5}+\frac{f^{\nu}}{\sqrt{1-f^{2}}}\gamma^{2}\gamma_{\nu}
\right)\psi(-t,-\vec{x}),
\end{eqnarray}
which has the desired properties.
Moreover, if the action of the new operators \textbf{TP} on the creation and
annihilation operators are 
\begin{eqnarray}
\textbf{TP}a^{s}_{\vec{p}}\textbf{PT} & = & a^{-s}_{\vec{p}} \\
\textbf{TP}b^{s}_{\vec{p}}\textbf{PT} & = & b^{-s}_{\vec{p}}
\end{eqnarray}
(where the $-s$ indicates a reversal of all the spin projections),
the action of these operators on the Dirac field in momentum space is 
\begin{equation}
\label{eq-modes}
\textbf{TP}\psi\textbf{PT}=\int\frac{d^{3}p}{(2\pi)^{3}}\frac{1}{\sqrt{E_{\vec{p}}}}
\left[a^{-s}_{\vec{p}}u^{s}(\vec{p}\,)^{*}e^{ipx}+(b^{-s}_{p})^{\dagger}v^{s}(\vec{p}\,)^{*}e^{-ipx}\right],
\end{equation}
where $u^{s}(\vec{p}\,)$ is the plane wave fermion solution of the modified Dirac equation.
It can be shown with some algebra that  
\begin{equation}
 (A^{-1})^{*}\gamma^{0}\gamma^{1}\gamma^{3}A u^{-s}(\vec{p}\,)= u^{s}(\vec{p}\,)^{*},
\end{equation}
and a similar relation also holds for antiparticle spinors $v^{s}(p)$.
(This generalizes the relations that holds for the eigenspinors in the usual, Lorentz-invariant
Dirac theory.) So the preceding formulas show the invariance of the
$f$ theory under physical inversion of all the spacetime coordinates.

\section{Discussion}

In this paper, we have demonstrated a number of useful facts and formulas describing
the SME fermion sector in the presence of a $f$ term, based on the relationship between such a theory and
a similar one with a $c$ term instead. One
important result was that the field redefinition that interchanges the two types of Lorentz violation
does not produce any extra effective potential terms in the Lagrangian, which means that the formula
(\ref{eq-ceff}) for $c_{{\rm eff}}$ is correct in even quantum theory.
We previously noticed how the effective $\beta$-function for
the nonlinear combination of coefficients $c^{\mu\nu}-\frac{1}{2}f^{\mu}f^{\nu}$ is free
of ambiguities up to second order in $f$. Since the equivalence of $c$ and $f$ is not anomalous
at the quantum level, we expect that the generalized quantity (\ref{eq-ceff}) will continue to
exhibit a scheme-independent RG scale dependence at all orders in $f$. Explicit calculations at
the higher orders should be possible, facilitated by the modified identities given in
section~\ref{sec-identities}.


Yet in spite of these results,
there are still potentially interesting technical questions about the equivalence between the 
$c$ and $f$ theories remaining to be answered.
We have shown how the mode expansion of the fermion field in the $f$ theory is affected by
the $f$ theory's \textbf{PT} operators (\ref{eq-modes}), but we did not explicitly show how this worked out
with the \textbf{C}, \textbf{P}, or \textbf{R}$_{j}$ (reflection) operators. In fact, we cannot
naively assume the transformation of the creation and annihilation operators under the action of
\textbf{C}, \textbf{P}, or \textbf{T} to be exactly the same as in a Lorentz-invariant free field theory.
To see the issue that arises, suppose that we begin in a theory with only a $c$, in which we may assume
that the $\textbf{C}'$, $\textbf{P}'$, and $\textbf{T}'$ transformations of fermion and antifermion creation
and annihilation should be the same as in the standard field theory. Then we perform the spinor transformation
to obtain instead an equivalent theory with a $f$ term.
However, the spinor redefinitions are not straightforward symmetry transformations, because they are not
generally unitary. This means that the
Fock spaces need not be same in the old and new theories. In short, the creation and annihilation operators
possibly will have some complicated nonunitary reshuffling of their eigenspinor coefficients, and so a
direct mode expansion of the field to see the effect of the \textbf{C} operator or of the \textbf{P}
operator alone is potentially nontrivial. Of course, nonunitary implementations of symmetry transformations
are nothing new in relativistic quantum field theory; because the Lorentz group in not compact, boost
transformations acting on spinors are also represented by nonunitary matrices. However, understanding
the relationships between the state spaces in the SME $c$ and $f$ theories is an interesting further
problem to be addressed.

\end{document}